\documentclass[twocolumn,prl,superscriptaddress,floatfix]{revtex4}
\usepackage{graphicx}
\usepackage{xcolor}
\usepackage{epsfig}
\usepackage{rotating}
\usepackage{amsmath}
\usepackage{amsfonts}
\usepackage{amssymb}
\usepackage{enumerate}
\usepackage{longtable}
\usepackage{appendix}
\usepackage{dcolumn}% Align table columns on decimal point
\usepackage{bm}

\usepackage{hyperref}\hypersetup{
    colorlinks=true,
    linkcolor=blue,
    filecolor=magenta,      
    urlcolor=blue,
    citecolor=blue
    }

\usepackage{ulem}

\usepackage{braket}

\begin{document}

\title{Instabilities of spin-1 Kitaev spin liquid phase in presence of single-ion anisotropies}

\author{Owen Bradley}
\affiliation{Department of Physics, University of California Davis, 
California 95616, USA}

\author{Rajiv R. P. Singh}
\affiliation{Department of Physics, University of California Davis, 
California 95616, USA}

\date{\rm\today}

\begin{abstract}
We study the spin-one Kitaev model on the honeycomb lattice in the presence of single-ion anisotropies. We consider two types of single ion anisotropies: A $D_{111}$ anisotropy which preserves the symmetry between $X$, $Y$, and $Z$ bonds but violates flux conservation and a $D_{100}$ anisotropy that breaks the symmetry between $X$, $Y$, and $Z$ bonds but preserves flux conservation. We use series expansion methods, degenerate perturbation theory, and exact diagonalization to study these systems. Large positive $D_{111}$ anisotropy leads to a simple product ground state with conventional magnon-like excitations, while large negative $D_{111}$ leads to a broken symmetry and degenerate ground states. For both signs there is a phase transition at a small $|D_{111}|\approx 0.12$ separating the more conventional phases from the Kitaev spin liquid phase. With large $D_{100}$ anisotropy, the ground state is a simple product state, but the model lacks conventional dispersive excitations due to the large number of conservation laws. Large negative $D_{100}$ leads to decoupled one-dimensional systems and many degenerate ground states. No evidence of a phase transition is seen in our numerical studies at any finite $D_{100}$. Convergence of the series expansion extrapolations all the way to $D_{100}=0$ suggests that the nontrivial Kitaev spin-liquid  is a singular limit of this type of single-ion anisotropy going to zero, which also restores symmetry between the $X$, $Y$, and $Z$ bonds.
\end{abstract}

%\pacs{74.70.-b,75.10.Jm,75.40.Gb,75.30.Ds}

\maketitle

%\section{I. Introduction}

\textit{Introduction.} Kitaev's spin-half honeycomb lattice model \cite{kitaev} provides a remarkable example of an exactly soluble emergent behavior with a quantum spin liquid ground state and Majorana fermion excitations
\cite{toric-code,balents-nature, balents2,baskaran1,vidal,feng,nussinov1,nussinov2,knolle14,knolle16,pollmann17,pollmann18}.
The search for such quantum spin liquid phases in spin-half materials remains a major focus of current research \cite{khaliullin, matsuda, trebst,kitaevm1,kitaevm2,banerjee,baharami,perkins}. 
Larger spin Kitaev models share some exotic properties of the spin-half models, namely they have conserved fluxes through each hexagon and no spin-spin correlations beyond nearest neighbors \cite{baskaran, koga1, koga2}. Yet, they are different in other key respects. 
As first proposed by Baskaran, Sen and Shankar \cite{baskaran} integer spin systems are unlikely to have Majorana fermions. The difference between integer and half integer spins is also highlighted in the work of Minakawa \textit{et al.} \cite{koga}, who found that introducing large anisotropy between $X$, $Y$, and $Z$ bonds leads to a very different type of ground state in integer spin systems with no long-range entanglement as compared with half-integer spin systems where similar anisotropy maps on to the well known Toric code model \cite{toric-code}. Numerical studies have found further evidence of a gap in the excitation spectra for integer spins and for field induced spin-liquid phases \cite{koga1, koga2, hykee1, hykee2, hykee3, sheng1, sheng2, ybk, nandini, qiang, you, hylee} as well as of large nearly degenerate subspaces giving rise to entropy plateaus \cite{koga1, koga2, oitmaa, bradley}. In a very recent paper, Chen \textit{et al.} \cite{chen-ybk} have shown the existence of emergent $Z_2$ spin liquid phase in the spin-one system with exotic deconfined anyonic excitations which are not Majorana fermions.

\begin{figure}[t!]
\centering
\includegraphics[width=0.9\columnwidth]{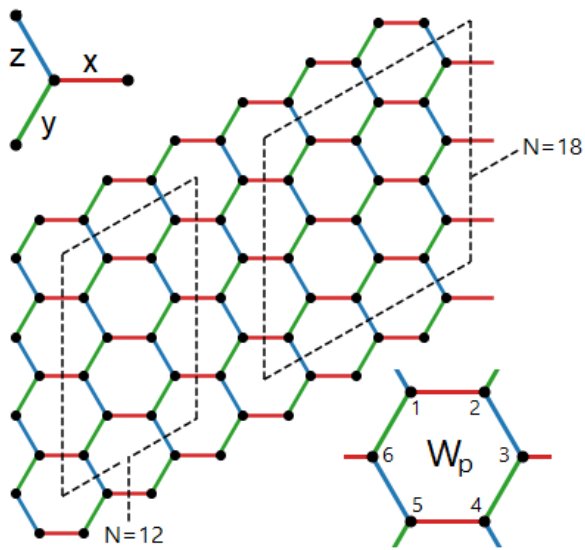}
\caption{Geometry of the honeycomb lattice, with the $x$, $y$, and $z$ bond directions indicated. The $N=12$ and $N=18$ site clusters studied using exact diagonalization are shown within dashed lines (with periodic boundary conditions). For each hexagonal plaquette (with sites labeled $1,\ldots,6$ as shown), one can define the flux operator $W_p$ given by Eq.~\eqref{FluxDef}.}
\label{lattice} 
\end{figure}

In this work we study the spin-one Kitaev model with two different types of single-ion anisotropies. The first model is given by:
\begin{equation}
\mathcal{H}_1 = \mathcal{H}_K + {\frac{D_{111}}{3}} \sum_{i} (S^x_i + S^y_i + S^z_i)^2,
\end{equation}
while the second model is:
\begin{equation}
\mathcal{H}_2 = \mathcal{H}_K + D_{100} \sum_{i} (S^z_i)^2,
\end{equation}
where $\mathcal{H}_k$ is the pure spin-one Kitaev honeycomb model Hamiltonian given by
\begin{equation}
\mathcal{H}_K = K \left(\sum_{\langle i,j \rangle} S^x_i S^x_j + \sum_{(i,k)} S^y_i S^y_k + \sum_{[i,l]} S^z_i S^z_l\right).
\end{equation}
Here the $X$, $Y$, and $Z$ couplings are on nearest neighbors of the honeycomb lattice pointing along the three sets of bond directions (see Fig.~\ref{lattice}).

It is evident that $D_{111}$ preserves the symmetry between $X$, $Y$, and $Z$ bonds whereas $D_{100}$ does not. 
For each hexagon in the lattice (with sites labeled $1, \ldots, 6$ as shown in Fig.~\ref{lattice}) one can define the plaquette flux operator 
\begin{equation}
W_p = e^{i \pi (S_1^z + S_2^y + S_3^x + S_4^z + S_5^y + S_6^x)}. \label{FluxDef}
\end{equation}
As shown in Ref.~\cite{baskaran}, the $W_p$ operators both commute with the Kitaev couplings and each other and have eigenvalues equal to $\pm 1$. Hence the model, in the absence of single-ion anisotropy, has conserved $Z_2$ flux variables on each hexagonal plaquette of the honeycomb lattice. 
One can show that $D_{100}$ term commutes with all the flux variables whereas $D_{111}$ term does not.
%In a system with full symmetry of the undistorted honeycomb lattice $D_{111}$ anisotropy should be present but not $D_{100}$.

For either type of anisotropy, large positive $D$ leads to a simple product ground state that can be studied by non-degenerate perturbation theory and high order series expansions. For large negative $D$, one can study the system by degenerate perturbation theory. For $D_{111}$ the phases at large positive or negative $D_{111}$ are conventional phases. We find in our numerical studies that these phases are separated from $D_{111}=0$ by phase transitions. 
%The transition can be seen as a sharp change in the entanglement entropy and the on-site state occupation probabilities. 
However, no such transition is evident with $D_{100}$ anisotropy. In this case even though the large $|D_{100}|$ ground states lack long-range entanglement, the phases remain exotic, either characterized by absence of conventional dispersive excitations or by a large number of ground states. Our study suggests that any long-range entangled quantum spin-liquid ground state depends crucially on the $D_{100}$ anisotropy going to zero. 

A nonzero $D_{100}$ cannot arise in a system with full symmetry of the honeycomb lattice where $X$, $Y$, and $Z$ bonds are equivalent but $D_{111}$ must always be present. Our study implies that experimental realizations of a Kitaev spin-liquid phase are possible in an undistorted honeycomb structure with $D_{111}$ single ion anisotropy, up to some moderate value of either sign. However, lattice distortions which allow $D_{100}$ terms to arise may immediately destabilize any phase with long range entanglement.

%\section{II. Model and perturbation theories}

\textit{Model and Perturbation Theories.} It is convenient to work in the $\ket{x}$, $\ket{y}$, $\ket{z}$ basis introduced by Koga \textit{et al.} \cite{koga1}, which can be expressed in terms of the $S^z$ basis as follows:
\begin{align}
\ket{x} &= -\frac{1}{\sqrt{2}}\left(\ket{m_s=1} - \ket{m_s=-1}\right)\\
\ket{y} &= \frac{i}{\sqrt{2}}\left(\ket{m_s=1} + \ket{m_s=-1}\right)\\
\ket{z} &= \ket{m_s=0}.
\end{align}

In this basis the spin operators are given by
\begin{equation}
    S^\alpha\ket{\beta} = i \epsilon_{\alpha \beta \gamma} \ket{\gamma}.
\end{equation}

The ground state at large positive $D_{111}$ is given by 
\begin{equation}
    \ket{\psi_g}=\prod_{i} \ket{0_i},
\end{equation}
where the state $\ket{0}$ at a site is given by
\begin{equation}
\ket{0} =  \frac{1}{\sqrt{3}} (\ket{x} + \ket{y} + \ket{z}),
\end{equation}
i.e.~the eigenvector of the $3 \times 3$ matrix $(S^x + S^y + S^z)^2$ with an eigenvalue of zero. To study this anisotropy we construct two states orthogonal to $\ket{0}$. In particular, we choose the states 
\begin{equation}
\ket{1} = \frac{1}{\sqrt{2}} (\ket{x} - \ket{y}),
\end{equation}
and, 
\begin{equation}
\ket{2}= \frac{1}{\sqrt{6}}  (\ket{x} + \ket{y} - 2\ket{z}). 
\end{equation}
The single-ion anisotropies are diagonal in this basis as are the flux variables. 
%The Kitaev couplings are given explicitly in the appendix. 

For large positive $D_{111}$, ground state properties can be obtained by nondegenerate perturbation theory which can be calculated by the linked-cluster method \cite{oitmaa_book,gelfand1,gelfand2}. The linked-cluster method states that a ground state property per site, $P$, can be expanded as a sum over all linked clusters $c$ as
\begin{equation}
    P=\sum_{c} L(c)\times W_P(c),
    \label{psum}
\end{equation}
where $L(c)$, called the lattice constant, is the number of ways the linked-cluster $c$ can be embedded in the lattice per lattice site. The quantity $W_P(c)$, called the weight of the cluster associated with the property $P$, is defined entirely by the property on the cluster and on its sub-clusters $s$ that can be embedded in $c$. It is defined as
\begin{equation}
    W_P(c)= N_c P(c) -\sum_{s\subset c} W_P(s),
\end{equation}
where $P(c)$ is the property calculated for the finite cluster and $N_c$ is number of sites in the cluster. One can show that the weight of a cluster with $N_b$ bonds only contributes in order $N_b$ or higher. Thus including all clusters with up to $N$ bonds in Eq.~\ref{psum} guarantees that one has the correct expansion in the thermodynamic limit to order $N$.

\begin{figure*}[htb!]
\centering
\includegraphics[width=1.65\columnwidth]{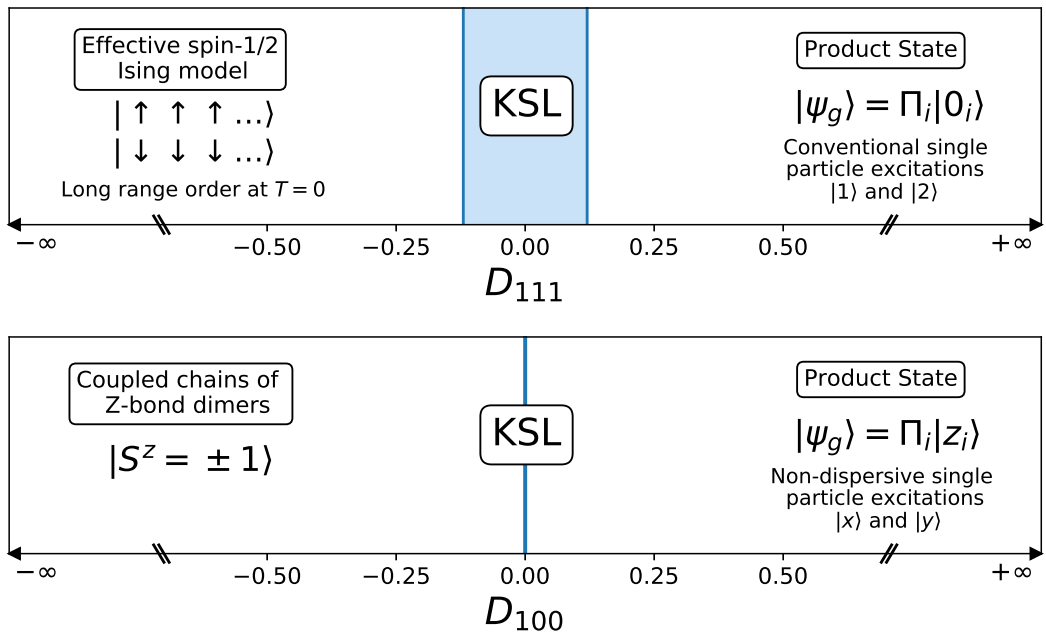}
\caption{Phase diagram of the spin-one Kitaev model in the presence of $D_{111}$ (top panel) and $D_{100}$ (bottom panel) single-ion anisotropy. The ground states observed in the limits of both large positive and large negative anisotropy, and the intermediate Kitaev spin liquid (KSL) region are indicated.}
\label{phase_diagram} 
\end{figure*}

For the expansion around the large positive $D_{111}$ we work in the basis of direct product of states $\ket{0}$, $\ket{1}$ and $\ket{2}$. In this basis the $D_{111}$ term is diagonal. It is useful to predetermine the $9\times 9$ perturbation matrix 
for the Kitaev couplings in the product basis of two sites. Once the matrix elements of the unperturbed Hamiltonian and the perturbation are known, perturbation theory for a finite system is reduced to simple recursion relations \cite{oitmaa_book,gelfand1,gelfand2}, which can be carried out through automated computer programs.

We obtain the ground state energy as
\begin{equation}
    E_g/D_{111} = \sum_n a_n (K/D_{111})^{n}.
\end{equation}
The occupation probability of the single-spin excited states $\ket{1}$ or $\ket{2}$ at a site in the ground state is given by
\begin{equation}
    n_{12} = \sum_n b_n (K/D_{111})^{n}.
\label{n12}
\end{equation}
The coefficients $a_n$ and $b_n$ up to $n=10$ are given in the Supplemental Material \cite{supplemental}. Numerical results will be presented in the next section when we compare with exact diagonalization.

For $K=0$ there are $2N$ single particle excitations corresponding to state $\ket{1}$ or $\ket{2}$ on a site. It is straightforward to construct the leading order in $K$ tight-binding hopping model for these excitations. 
%It is also given in the appendix. 
The system clearly has conventional single-particle excitations.
%with a minimum at the $\Gamma$ point in the Brillouin Zone.

%\begin{table}
%\begin{center}
% \begin{tabular}{c c c} 
% \hline\hline
% n & $c_n$ & $d_n$ \\
% \hline
%           1 & -0.500000000000000   &    0.500000000000000 \\
%           2 & 0.375000000000000    &   -1.12500000000000  \\  
%           3 & -0.625000000000000   &     3.12500000000000 \\  
%           4 & 1.36132812500000     &  -9.52929687500000   \\ 
%           5 & -3.40995279948253    &   30.6895751953348   \\
%           6 &  9.31669814497736    &   -102.483679594831  \\  
%           7 & -27.0160930127270    &    351.209209165077  \\  
%           8 & 81.8162037114470     &   -1227.24305566977    \\ 
%           9 & -256.114609164988    &    4353.94835580449  \\   
%          10 &  822.924780762607    &   -15635.5708344931  \\  
%          11 & -2700.51360857928    &   56710.7857801259   \\ 
%          12 &  9017.61445118801    &   -207405.132377063  \\
%\hline\hline
%\end{tabular}
%\end{center}
%\caption{Series coefficients for ground state energy $E$ and %occupation probabilities $n_{xy}$ for the case of $D_{100}$ %anisotropy.}
%\end{table}

For large negative $D_{111}$ the states $\ket{1}$ and $\ket{2}$ provide degenerate on-site ground states. In this $2^N$ dimensional Hilbert space one can obtain the effective Hamiltonian by degenerate perturbation theory. Remarkably, in this reduced subspace, $S^x$, $S^y$, and $S^z$ become identical off-diagonal operators and the system maps on to an effective spin-half Ising model, with commuting terms, that has two degenerate ground states with long-range order. 
%These results are confirmed by exact diagonalization.

For the case of large $D_{100}$ the ground state is given by
\begin{equation}
    \ket{\psi_g}=\prod_{i} \ket{z_i}.
\end{equation}
One can study its properties by non-degenerate perturbation theory using the linked-cluster method \cite{oitmaa_book,gelfand1,gelfand2}. The ground state energy series is
\begin{equation}
    E_g/D_{100} = \sum_n c_n (K/D_{100})^{2n}.
\end{equation}
This model is invariant under a change of sign of the Kitaev couplings and hence the properties depend only on $(K/D_{100})^2$.

The occupation probability of the single-spin excited states $\ket{x}$ or $\ket{y}$ at a site in the ground state is given by
\begin{equation}
    n_{xy} = \sum_n d_n (K/D_{100})^{2n}.
\label{nxy}
\end{equation}
The coefficients $c_n$ and $d_n$ up to $n=12$ are given in the Supplemental Material \cite{supplemental}. Numerical results will be presented in the next section when we compare with exact diagonalization.

Note that despite the product ground state, this system remains unconventional. Due to various conservation laws, single-particle states remain confined to single bonds, an $\ket{x}$ excitation is confined to a single $Y$ bond, where as a $\ket{y}$ excitation is confined to a single $X$ bond. Only states in the zero-flux sector can be delocalized \cite{sen2}.
%and there are no conventional dispersive quasiparticles.

At large negative $D_{100}$, we need to carry out a degenerate perturbation theory in the space of states $\ket{x}$ and $\ket{y}$ on the different sites. In this case, it is easier to go back to the $S^z$ basis. In the degenerate $2^N$ dimensional Hilbert space given by $\ket{S^z=\pm 1}$, the system at first breaks into decoupled dimers along the $Z$ bonds. Depending on the sign of the Kitaev couplings, in first-order perturbation theory, the lower energy state corresponds to parallel or antiparallel spins on each dimer. This still leaves $2^{N/2}$ degenerate states. A higher order degenerate perturbation theory in this subspace is needed. In the $4^{th}$ order, the system breaks into coupled chains of $Z$-bond dimers. The $Z$-bond dimers in a row are coupled by a transverse Ising exchange coupling between effective spin-half degrees of freedom on neighboring dimers. Thus, there are two degenerate ground states for each such chain of $Z$-bond dimers
and the system has large but nonextensive ground state degeneracy. In Fig.~\ref{phase_diagram} we show a phase diagram illustrating the ground states observed for both $D_{111}$ and $D_{100}$ anisotropy.

To study the model near $D=0$ it is essential to perform numerical studies.

\begin{figure}[t!]
\centering
\includegraphics[width=\columnwidth]{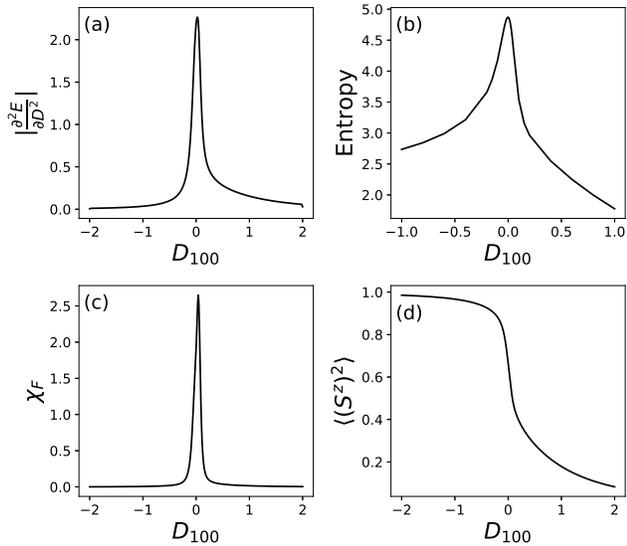}
\caption{Exact diagonalization results for a $N=18$ site cluster with $D_{100}$ anisotropy, with $|K|=1$. We show (a) the second derivative of ground state energy (arbitrary units), (b) entanglement entropy, (c) fidelity susceptibility (arbitrary units), and (d) $\langle (S^z)^2 \rangle$ as a function of $D_{100}$. Note that $\langle (S^z)^2 \rangle$ is equivalent to $n_{xy}$ as defined in Eq.~\eqref{nxy}.}
\label{ed_d100_4panel} 
\end{figure}
 
\begin{figure}[ht!]
\centering
\includegraphics[width=\columnwidth]{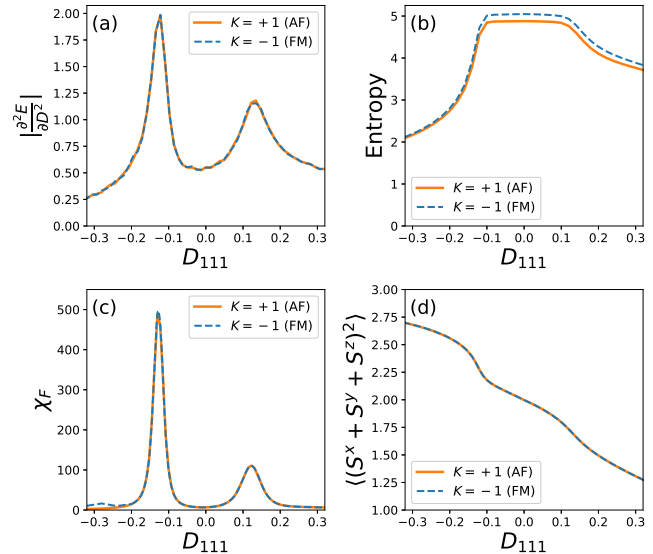}
\caption{Exact diagonalization results for a $N=18$ site cluster with $D_{111}$ anisotropy, for both ferromagnetic (FM) and antiferromagnetic (AF) Kitaev couplings. We show (a) the second derivative of ground state energy (arbitrary units), (b) entanglement entropy, (c) fidelity susceptibility (arbitrary units), and (d) $\langle (S^x + S^y + S^z)^2 \rangle$ as a function of $D_{111}$. Note that $\frac{1}{3}\langle (S^x + S^y + S^z)^2 \rangle$ is equivalent to $n_{12}$ as defined in Eq.~\eqref{n12}.}
\label{ed_d111_4panel}
\end{figure}

%\section{III. Numerical Studies}

%For each hexagon in the lattice (with sites labeled $1, \ldots, 6$ as shown in Fig.~\ref{lattice}) one can define the plaquette flux operator 
%\begin{equation}
%W_p = e^{i \pi (S_1^z + S_2^y + S_3^x + S_4^z + S_5^y + S_6^x)}, \label{FluxDef}
%\end{equation}
%As shown in Ref.~\cite{baskaran}, the $W_p$ operators both commute with the Kitaev couplings and each other and have eigenvalues equal to $\pm 1$. Hence the model in the absence of single-ion anisotropy has conserved $Z_2$ flux variables on each hexagonal plaquette of the honeycomb lattice.

\textit{Numerical Studies.} We study the ground states of the model with different values of the anisotropy using Lanczos exact diagonalization of 12 and 18 site clusters for both $D_{111}$ and $D_{100}$ anisotropy. The larger system size study is enabled in the latter case by the conserved fluxes, which reduce the connected Hilbert space size, and hence the ground state is always found in the zero flux sector. For $D_{111}$ anisotropy the fluxes are not conserved, however, the translational symmetries of the 18-site cluster (along with an inversion symmetry) give a reduced Hilbert space dimension $\sim \frac{3^{18}}{18}$, enabling Lanczos exact diagonalization of this larger system size.

We show below results of ground state energy and its second derivative, on-site occupation probabilities, entanglement entropy when the system is divided into two equal halves, and fidelity susceptibility defined as
\begin{equation}
    \chi_F = \frac{2\left[1 -|\braket{\psi_g(x)|\psi_g(x+dx)}|\right]}{dx^2}.
\end{equation}

In Fig.~\ref{ed_d100_4panel} the results for various ground state properties with the $D_{100}$ anisotropy from Lanczos diagonalization of the 18-site cluster are shown. In Fig.~\ref{ed_d111_4panel} the results are shown for the corresponding study of the 18-site cluster with $D_{111}$ anisotropy. 

\begin{figure}[t!]
\centering
\includegraphics[width=\columnwidth]{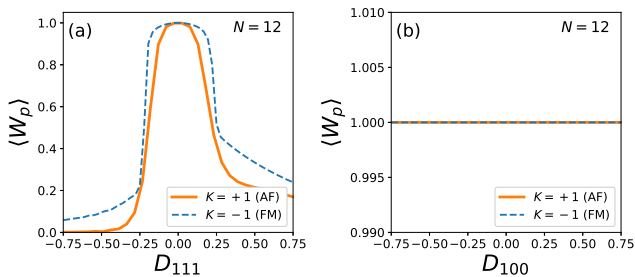}
\caption{Average value of the plaquette flux operator $W_p$ as a function of (a) $D_{111}$ anisotropy and (b) $D_{100}$ anisotropy, for both ferromagnetic (FM) and antiferromagnetic (AF) Kitaev couplings. Exact diagonalization results are shown for a $N=12$ site cluster.}
\label{flux} 
\end{figure}
 
It is evident from the figures that the $D_{111}$ model undergoes a phase transition as the $D_{111}=0$ limit is approached. For the 18-site cluster we find no significant difference in behavior for ferromagnetic and antiferromagnetic Kitaev couplings. Peaks in the second derivative of the energy and fidelity susceptibility occur at $D_{111} \approx \pm 0.12$, along with a region of maximum entanglement entropy between these values. The value of the anisotropy at which the transition occurs is similar in the two cases. The average plaquette flux $\langle W_p \rangle$ approaches 1 as the $D_{111}=0$ limit is approached as expected, changing rapidly in the transition region and falling to zero in the limit of large negative or large positive $D_{111}$ anisotropy, as shown in Fig.~\ref{flux}(a). 

In contrast, for the $D_{100}$ anisotropy the sharpest changes occur at $D_{100}=0$. The entanglement entropy, fidelity susceptibility, and second derivative of ground state energy are all sharply peaked very near $D_{100}=0$. In the finite system the peaks are not strictly at $D_{100}=0$, but they are also system size dependent and consistent with the singularity being right at $D_{100}=0$. The transition at $D_{100}=0$ is further supported by comparison with the series analysis, which is done in the thermodynamic limit, presented in the next section. Since the ground state is always in the zero flux sector, we have $\langle W_p \rangle = 1$ for all values of $D_{100}$ as shown in Fig.~\ref{flux}(b).

\begin{figure}[t!]
\centering
\includegraphics[width=\columnwidth]{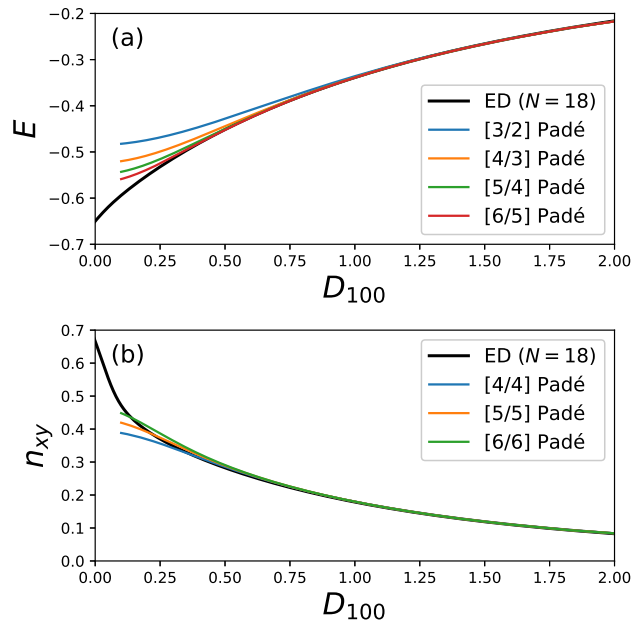}
\caption{(a) Ground state energy per site and (b) local occupation of excited states as a function of $D_{100}$. Exact diagonalization results are shown for a $N=18$ site cluster, along with Pad\'e approximants to each series expansion.}
 \label{energy_and_nxy} \end{figure}

\begin{figure}[h!]
\centering
\includegraphics[width=\columnwidth]{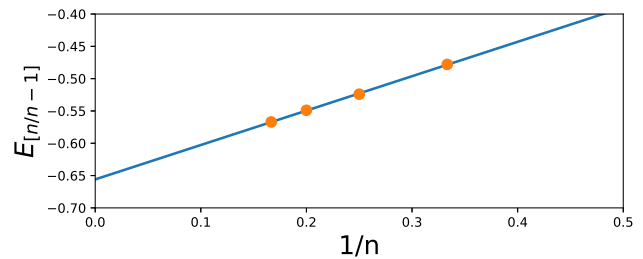}
\caption{Asymptotic $K/D_{100} \to \infty$ value of ground state energy from $[n/n-1]$ Pad\'e approximant for the ground state energy series is further extrapolated as a function of $1/n$ to get an estimate for the $D_{100}=0$ ground state energy. It is found to be approximately $E/K=-0.656$.}
\label{energy-extrapolation}
\end{figure} 

\begin{figure}[t!]
\centering
\includegraphics[width=\columnwidth]{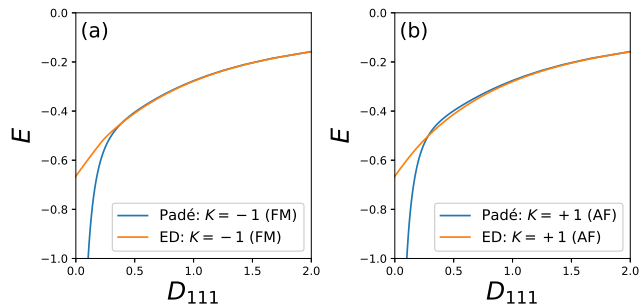}
\caption{Ground state energy per site as a function of $D_{111}$ anisotropy, for both (a) ferromagnetic and (b) antiferromagnetic Kitaev couplings. Results are shown comparing exact diagonalization data for a $N=12$ site cluster and a Pad\'e approximant of series expansion data.}
 \label{energy-111} \end{figure}

%\section{IV. Comparisons with series expansions and Discussion}
\textit{Comparison with Series Expansion and Discussion.} 
A direct comparison of the energy and state occupation $n_{xy}$ for positive $D_{100}$ are shown in Fig.~\ref{energy_and_nxy}.
For the $D_{100}$ anisotropy the series are in powers of $(K/D_{100})^2$. One can estimate the ground state energy in the large $K/D_{100}$ limit by using Pad\'e extrapolation. Since the energy in this limit must go as $K$, we first square the energy series. The resulting series are analyzed by $[n/n-1]$ Pad\'e approximants. This ensures the correct large $K/D_{100}$ behavior. The series results for different Pad\'e approximants are shown. One can see that the range of convergence is improving as more terms are added. However, the convergence slows down as $D_{100}$ goes to zero. The extrapolated values at $K/D_{100}\to \infty$ from $[n/n-1]$ Pad\'e are then further extrapolated as a function of $1/n$ in Fig~\ref{energy-extrapolation}. The linear fit to $1/n$ gives ground state energy at $D_{100}=0$ of $E/K=-0.656$, which is close to the value $E/K \sim -0.65$ obtained from previous numerical studies of finite-size clusters \cite{koga1, hykee2}. The success of this extrapolation is evidence that the transition to a long-range entangled state happens very close to $D_{100}=0$, likely right at $D_{100}=0$. Also, the need to extrapolate Pad\'e results with $1/n$ suggests that $D_{100}=0$ is a singular limit. Thus, for any nonzero $D_{100}>0$, the ground state is adiabatically connected to the product state at large $D_{100}$ and hence lacks long-range entanglement.

For the $D_{111}$ case also the series expansion converges well at large $D_{111}$ values until the transition region is reached beyond which the series expansion disagrees sharply with the exact diagonalization results, as shown in Fig.~\ref{energy-111}. No meaningful estimate of the properties in the $D_{111}\to 0$ limit can be obtained from the series. This is consistent with the existence of a phase transition in the model at finite $D_{111}$.

We note that within perturbation theory the ferromagnetic and antiferromagnetic Kitaev models are identical in first few orders. The difference first arises in the order $(K/D_{111})^7.$ Similarly, the leading effective Hamiltonian at large negative $D_{111}$ and the selection of order does not depend on the sign of Kitaev exchange couplings. Thus any difference between ferromagnetic and antiferromagnetic Kitaev couplings is a higher-order process and comparison with the exact diagonalization suggests that it is numerically not very significant. We note that the addition of other exchange couplings (such as Heisenberg and Gamma couplings) and magnetic fields to the Kitaev Hamiltonian causes sharp differences between ferromagnetic and antiferromagnetic Kitaev models in both the spin-half and spin-one cases \cite{trebst,hykee1,hykee2,hykee3,sheng1,sheng2} and these deserve further attention in the presence of single-ion anisotropies.

%\section{V. Summary and Conclusions}
\textit{Summary and Conclusions.} In this paper we have studied the spin-one Kitaev model with two types of single-ion anisotropies using a variety of perturbative and numerical methods. We find that the $D_{111}$ anisotropy, which preserves the symmetry between $X$, $Y$, and $Z$ bonds but violates flux conservation, leads to conventional phases and excitations at large anisotropy values. There is a phase transition at a modest value of $D_{111}/K \approx 0.12$ that separates these conventional phases from the Kitaev spin liquid. In contrast, for large $D_{100}$ anisotropy, even though the ground states are simple and lack long-range entanglement the system remains exotic at both large negative and positive $D_{100}$ values. For positive $D_{100}$ there is a non-degenerate ground state but no conventional dispersive quasiparticle excitations. For negative $D_{100}$, the system decouples into chains of $Z$-bond dimers that are coupled by an effective Ising coupling along the chain. Thus there are two degenerate ground states for each such chain. No signature of a phase transition is seen in our study as a function of $D_{100}$. The fidelity, second derivative of free energy and entanglement entropy are all sharply peaked near $D_{100}=0$ suggesting that the Kitaev spin liquid emerges only upon $D_{100}$ going to zero and restoration of symmetry between $X$, $Y$, and $Z$ bonds. We emphasize that the Kitaev spin-liquid becomes immediately unstable in the presence of $D_{100}$ anisotropy.

Candidate materials for spin-one Kitaev spin liquids and underlying exchange mechanisms have been recently proposed \cite{hykee1}. Real spin-one materials are known to always have some single-ion anisotropies. A material with the full symmetry of the honeycomb lattice will likely have only $D_{111}$ anisotropy. Our work shows that the anisotropy value must be small compared to Kitaev couplings to realize a Kitaev spin liquid ground state. However, distortions which can allow nonzero $D_{100}$ anisotropy may be particularly destabilizing to the long-range entangled spin-liquid phase. 

In future, it may be useful to study the nature of the phase transitions for $D_{111}$ anisotropy. Density matrix renormalization group or tensor network studies on larger system sizes may be helpful in this respect \cite{chen-ybk}. For a positive sign of $D_{111}$ the two phases on either side of the transition have no broken symmetries, hence we speculate that the transition may be purely topological in nature, although the transition could be first order. It would also be interesting to better elucidate the mechanism for loss of long-range entanglement with $D_{100}$ anisotropy, which should also throw further light on the nature of the spin-liquid phase. 

\textit{Acknowledgments.~} This work is supported in part by National Science Foundation Grant No.~NSF-DMR 1855111.  

\bibliography{main}
\end{document}